\documentclass[a4paper]{article}
\usepackage{multirow}
\usepackage{ISCSLP2022}
\usepackage{float}
\usepackage{enumerate}
\usepackage{url}

\title{AccentSpeech: Learning Accent from Crowd-sourced Data for Target Speaker TTS with Accents}

\name{Yongmao Zhang$^1$, Zhichao Wang$^1$, Peiji Yang$^2$, Hongshen Sun$^2$, Zhisheng Wang$^2$, Lei Xie$^1$$^*$}
\address{
  $^1$Audio, Speech and Language Processing Group (ASLP@NPU)\\School of Computer Science,
  Northwestern Polytechnical University, Xi’an, China\\
  $^2$Tencent, Shenzhen, China}
\email{zym@mail.nwpu.edu.cn, zcwang\_aslp@mail.nwpu.edu.cn, lxie@nwpu.edu.cn, \\
\{peijiyang, seanhssun, plorywang\}@tencent.com}

\begin{document}

\maketitle
\begin{abstract}

\vspace{-2pt} 

Learning accent from crowd-sourced data is a feasible way to achieve a target speaker TTS system that can synthesize accent speech. To this end, there are two challenging problems to be solved. First, direct use of the poor acoustic quality crowd-sourced data and the target speaker data in accent transfer will apparently lead to synthetic speech with degraded quality. To mitigate this problem, we take a bottleneck feature (BN) based TTS approach, in which TTS is decomposed into a Text-to-BN (T2BN) module to learn accent and a BN-to-Mel (BN2Mel) module to learn speaker timbre, where neural network based BN feature serves as the intermediate representation that are robust to noise interference. Second, direct training T2BN using the crowd-sourced data in the two-stage system will produce accent speech of target speaker with poor prosody. This is because the the crowd-sourced recordings are contributed from the ordinary unprofessional speakers. To tackle this problem, we update the two-stage approach to a novel three-stage approach, where T2BN and BN2Mel are trained using the high-quality target speaker data and a new BN-to-BN module is plugged in between the two modules to perform accent transfer. To train the BN2BN module, the parallel unaccented and accented BN features are obtained by a proposed data augmentation 
procedure. Finally the proposed three-stage approach manages to produce accent speech for the target speaker with good prosody, as the prosody pattern is inherited from the professional target speaker and accent transfer is achieved by the BN2BN module at the same time. The proposed approach, named as \textit{AccentSpeech}, is validated in a Mandarin TTS accent transfer task.

\end{abstract}

\renewcommand{\thefootnote}{\fnsymbol{footnote}}
\footnotetext{* Corresponding author.}

\vspace{-6pt} 
\section{Introduction}

\vspace{-0.1cm} 
With the help of deep learning, corpus based speech synthesis has made significant progress in both quality and naturalness in recent years~\cite{DBLP:journals/corr/abs-2106-15561}. Besides delivering linguistic information, para-linguistic aspects of speech, such as speaking style~\cite{DBLP:conf/icml/WangSZRBSXJRS18,DBLP:conf/icassp/ZhangPHL19,DBLP:conf/asru/AnWYMX19}, emotion~\cite{DBLP:conf/iscslp/LiYXX21, DBLP:conf/slt/LeiYX21} as well as accent, are highly desired in text-to-speech (TTS) systems. 

\textit{Accent} is a distinctive mode of pronunciation of a language, especially one associated with a particular nation, locality, or social class. Early studies have attempted to convert non-native accent speech to native accent speech for language learning~\cite{DBLP:conf/icassp/AryalG14a} or speech communication~\cite{DBLP:journals/speech/FelpsBG09, DBLP:conf/icassp/ZhaoSLCG18, DBLP:conf/icassp/LiuWCSWKWLSYM20}. It is also ideal for a speech generation task, such as TTS, to provide synthetic speech of a target speaker with multiple accents. However, this is not a trivial problem as it is difficult to find a professional speaker who can speak various types of accents. \textit{Accent transfer} thus becomes a feasible solution, where synthesizing target speaker's accent speech is realized by transferring the desired accent from another speaker. In practice, it is much easier to collect crowd-sourced accent speech data. Therefore in this paper, we aim to build an accent TTS system named \textit{AccentSpeech}, which learns various accents from crowd-sourced data. To this end, there are least two problems have to be addressed when transferring the accent from the crowd-sourced data to the target speaker.

The first problem is \textit{poor acoustic quality}. As we know, crowd-sourced data, usually recorded using personal on-device microphone in an ordinary room, may inevitably contain background noise, reverberation and channel distortion. Using such poor quality data in neural acoustic modeling and vocoding will apparently lead to poor quality synthetic speech and artifacts.

To mitigate this problem, we take a bottleneck feature (BN) based TTS approach, in which text-to-speech is decomposed into a Text-to-BN (T2BN) module and a BN-to-Mel (BN2Mel) module, where neural network based bottleneck feature serves as the intermediate representation. Bottleneck feature or phonetic posteriorgram (PPG) was originally used in voice conversion within a recognition-synthesis framework~\cite{DBLP:conf/icmcs/SunLWKM16}, where a massive data pre-trained speaker-independent speech recognition model is first adopted to extract PPG or BN feature from source speech uttered by the source speaker, and then a voice conversion acoustic model is learnt to map the PPG/BN feature to the the target speaker’s speech. The BN feature, resulted from a well-trained neural ASR model, is considered to be robust to noise and other interference~\cite{DBLP:journals/taslp/LiDGH14}. Such a decomposition approach has been recently adopted in TTS task as well~\cite{DBLP:conf/icassp/LiuYSY22,DBLP:conf/icassp/DaiCCTLXTWW22}. With the help of BN features, the BN2Mel module aims to model target speaker's timbre with high-quality data while the T2BN module is used for modeling linguistic content, prosody/style and other information using massive multi-speaker data. Specifically, in our approach, the linguistic content and specific accent information are embedded in the BN feature extracted from an accent speaker while the BN2Mel module learns the timbre of the target speaker. Thanks to the noise-robust BN feature, the T2BN + BN2Mel process eventually produces high-quality target speaker synthetic speech with desired accent.

The second problem is \textit{poor speaker prosody}. The BN feature is considered to contain prosodic information, while its framewise nature makes it embed duration information. The above two-stage approach works well with the premise that the T2BN module is trained using data from a professional speaker with good speech prosody. However, the crowd-sourced speech data are from unprofessional ordinary speakers. The (bad) prosody pattern will be learned from these data when training the T2BN model, eventually leading to synthetic accent speech with unsatisfied prosody. 

To tackle this problem, we insert a BN2BN \textit{accent transfer} module in-between and result in our AccentSpeech system. Specifically, we first use target-speaker high-quality speech to train the T2BN module, the BN2Mel module and the vocoder. Then the text transcripts, associated with the speech from the accent speaker, are fed into the  T2BN module to generate the corresponding BN features (unaccented BN) which are considered to embed the linguistic information as well as the target speaker's original accent. Meanwhile, the same set of training speech from the accent speaker goes through the ASR model, resulting in the BN features (accented BN) which are regarded to embed the linguistic information as well as the accent speaker's accent information. We subsequently use a neural network based BN2BN accent transfer module to learn the accent conversion from the parallel unaccented BN - accented BN features. Finally the proposed T2BN + BN2BN + BN2Mel process manages to produce accent speech of the target speaker with good prosody, as the prosody pattern is actually inherited from the professional target speaker and accent transfer is achieved by the BN2BN module at the same time.

Experiments on a open-source professional speaker and a crowd-sourced multi-speaker accent dataset~\cite{DBLP:conf/nips/Tang0XSLZWTXZYL21} show the superiority of the proposed AccentSpeech. We mange to build a high-quality target speaker TTS system with various accents and natural prosody.

\begin{figure}[t]
  \centering
  \includegraphics[width=0.9\linewidth]{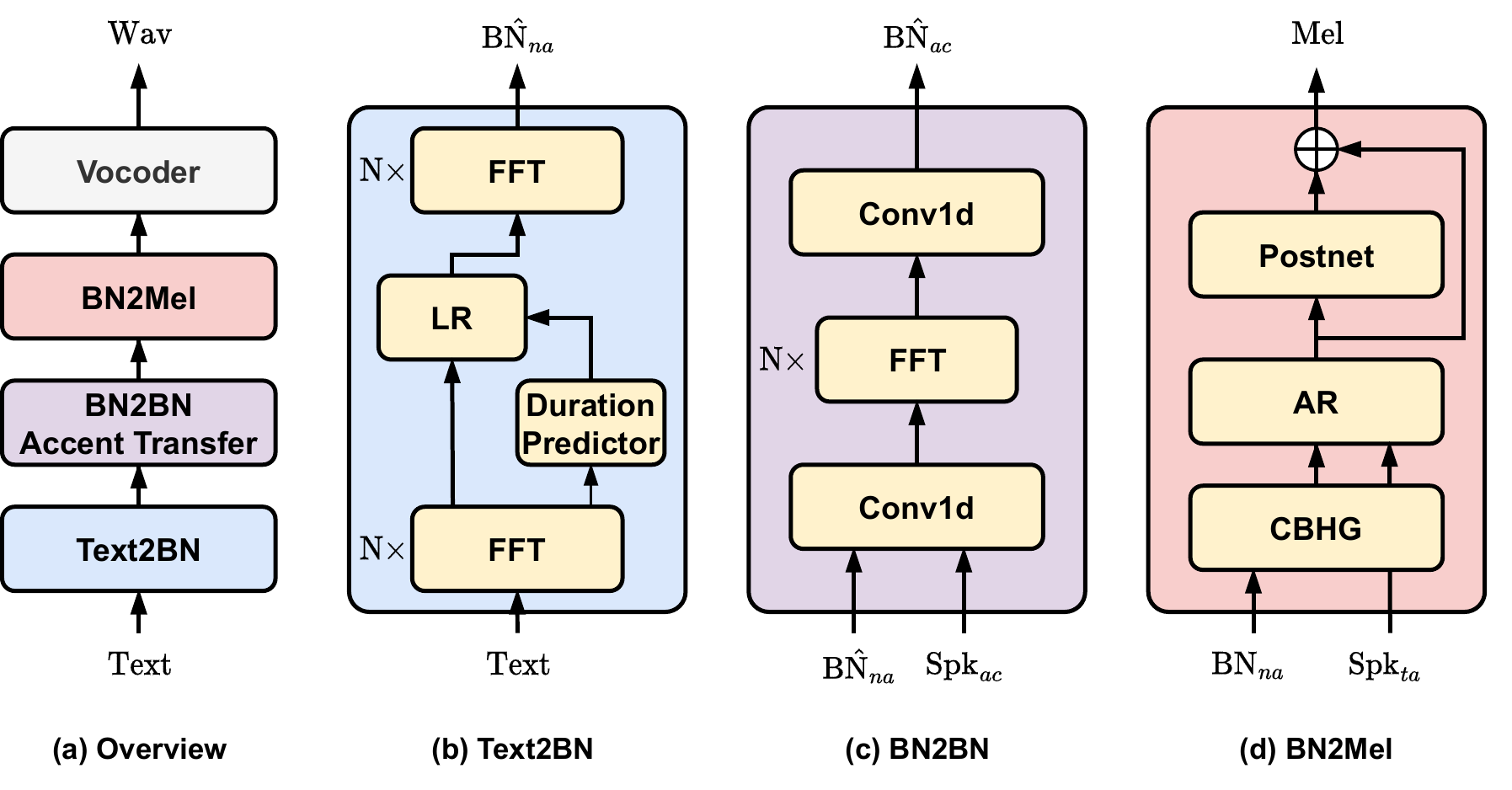}
  \vspace{-3pt} 
  \caption{The Overall architexture of AccentSpeech. A BN2BN model is trained separately for each accent, and the speakers in (c) have the same accent but express it differently, like accent intensity, in our dataset.}
  \label{fig:model}
  \vspace{-2em}
\end{figure}

\vspace{-10pt} 
\section{Method}

\vspace{-6pt} 
\subsection{System Overview}

\vspace{-0.1cm}

Suppose we have a target speaker with standard accent (unaccented -- $ua$) and parallel text-audio data. The corresponding BN features are extracted through an ASR system, resulting in $\text{BN}_{ua}$. Meanwile, we have an accented speaker with a specific accent $ac$ and parallel text-audio data collected in the crowd-sourced manner. Likewise, the corresponding BN features extracted through the same ASR system is denoted as $\text{BN}_{ac}$.
As shown in Figure~\ref{fig:model}, to decouple accent-related information from speaker timbre for further recombination, we design AccentSpeech in a cascade manner with three neural network modules: Text-to-BN~(T2BN), BN-to-BN~(BN2BN) and BN-to-Mel~(BN2Mel). 

First, the T2BN module, trained using the target speaker data, expands input text to unaccented BN ($\widehat{\text{BN}}_{ua}$), which is considered to contain linguistic and prosody information without speaker timbre. Similar to Fastspeech~\cite{DBLP:conf/nips/RenRTQZZL19}, the T2BN model consists of several feed-forward transformer blocks (FFT), length regulator (LR) and duration predictor. 
The BN2BN model aims to transform $\widehat{\text{BN}}_{ua}$ to the corresponding $\widehat{\text{BN}}_{ac}$ which is believed to have the target accent information ($ac$). It is composed of a set of FFT and convolution layers. Subsequently, the BN2Mel model takes $\widehat{\text{BN}}_{ac}$ as input and outputs the mel-spectrum with target speaker timbre and accent ($ac$).  For the BN2Mel model, we use the CBHG~\cite{DBLP:conf/icml/Skerry-RyanBXWS18} module as the encoder, and meanwhile, the decoder adopts an auto-regressive module and postnet~\cite{DBLP:conf/icassp/ShenPWSJYCZWRSA18}. With the generated spectrum, a HiFiGAN~~\cite{DBLP:conf/nips/KongKB20} vocoder is finally adopted to reconstruct the waveform. Note that the BN2Mel module and the vocoder are both trained using the target speaker data. The key to realize accent transfer is the BN2BN model which will be described in Section 2.3 in detail.

\vspace{-6pt} 
\subsection{Speaker-Accent Decomposition based on BN features}

\vspace{-0.1cm}
With the high cost of collecting speech recordings with different accents from each single speaker, it is necessary to decouple accent and speaker information from speech and recombine them to realize synthetic accented speech for each speaker. Previously, adversarial training~\cite{DBLP:journals/spl/YangWX20, DBLP:conf/interspeech/CongY0YW20} was used to achieve this goal. But taking into account the unstable process of adversarial training and unavailability of enough high-quality multi-accent data, we propose to use a cascade approach based on neural network bottleneck (BN) feature as a bridge. BN feature is usually the feature map of a neural network layer. Specifically, in our approach, BN is the output of the ASR encoder which is commonly believed to contain linguistic and prosodic information, such as pronunciation, intonation and accent, but limited speaker information~\cite{DBLP:journals/speech/FelpsBG09, DBLP:conf/icassp/ZhaoSLCG18, DBLP:conf/icmcs/SunLWKM16}.

The ASR model is usually trained with a large multi-speaker multi-condition dataset and the BN feature resulted from it is also believed to be speaker-independent and noise-robust. This inspired us to leverage crowd-sourced (noisy) accented speech data to realize accent transfer.

\begin{figure*}[t]
  \centering
  \includegraphics[width=0.9\linewidth]{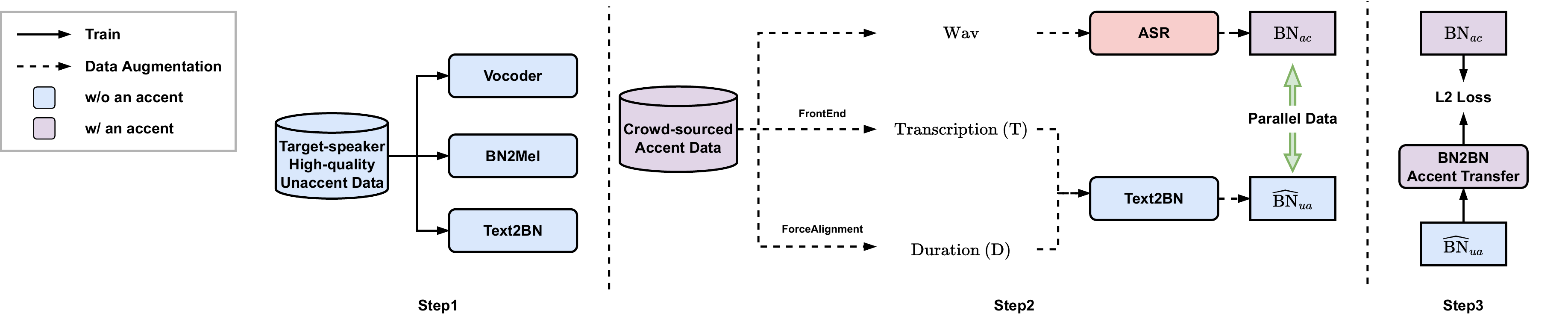}
  \vspace{-4pt}
  \caption{The training process of AccentSpeech.}
  \label{fig:data_expansion}
  \vspace{-20pt}
\end{figure*}

\vspace{-6pt} 
\subsection{Accent Transfer with Data Augmentation\label{accent transfer}}

\vspace{-0.1cm}

The frame-level BN feature is considered to contain prosodic information, including phoneme duration. If we directly use the crowd-sourced data to train the Text2BN model, poor speech prosody from the unprofessional speaker will be transferred to the synthetic speech of the professional target speaker along with the accent. To avoid learning poor speech prosody from the crowd-sourced data, we propose to add a BN2BN model in-between the Text2BN model and the BN2Mel model. Specifically, the BN2BN model aims to learn the mapping function between $\text{BN}_{ua}$ and $\text{BN}_{ac}$. However, we have different sets of data from the target speaker and the crowd-sourced accent speaker. To learn the mapping, we need frame-wise time-aligned parallel data between the two speakers. To this end, we use the data augmentation strategy shown in Figure~\ref{fig:data_expansion}.

First, we use the unaccented data from the target speaker to train the T2BN model, the BN2Mel model and the vocoder. The BN features are extracted using a pre-trained ASR model.

Second, with the speech recording and corresponding transcription $\text{T}$ obtained by text front-end of Mandarin from the accented speaker, the phoneme-level duration $\text{D}$ of each speech recording can be obtained by a force-alignment tool~\cite{povey2011kaldi}. And then, we use the T2BN model to generate $\widehat{\text{BN}}_{ua}$ which is in parallel with $\text{BN}_{ac}$ from the accented speaker. This process can be described as: 

\vspace{-0.3cm}
\begin{equation}
\widehat{\text{BN}}_{ua}=\text{T2BN}(\text{T},\text{D}).
\end{equation}
\vspace{-0.4cm}

Finally, the parallel data \{$\widehat{\text{BN}}_{ua},\text{BN}_{ac}$\} is available for training the BN2BN model for accent transfer. Here we assume that the duration is unrelated to the accent. The BN2BN model is trained to conduct frame-wise mapping between $\text{BN}_{ac}$ and $\widehat{\text{BN}}_{ua}$. This process can be described as

\vspace{-10pt}
\begin{equation}
\widehat{\text{BN}}_{ac}=\text{BN2BN}(\widehat{\text{BN}}_{ua},\text{Spk}_{ac}).
\end{equation}
\vspace{-14pt}

Note that in our implementation, we have multiple speakers speaking the same type of accent while each speaker may have different personal speaking style and accentedness (degree of accent). We use an extra accent speaker identity $\text{Spk}_{ac}$  as a condition for accent transfer.

\vspace{-0.2cm} 
\section{Experiments}

\vspace{-0.2cm} 
\subsection{Dataset}

\vspace{-0.1cm} 

The data for experimentation consists of high-quality standard Mandarin speech data and low-quality accented Mandarin speech data. Specifically, we use DB1~\cite{db1} as the target high-quality data which contains 10,000 utterances recorded in a studio from a professional female anchor with a total audio length of about 10 hours. We select the crowd-sourced data recorded from three specific cities -- Chengdu, Xi'an and Zhengzhou -- from the KeSpeech~\cite{DBLP:conf/nips/Tang0XSLZWTXZYL21} corpus as the accent data. The speech recordings from each city are collected from multiple speakers with clear local accents. There are 30,566, 32,323, and 21,901 utterances from Chengdu, Xi'an and Zhengzhou, respectively. The crowd-sourced data are recorded from mobile phones with typical reading style, ambient noise and reverberation.

All the audio recordings are downsampled to 16kHz. We use 80-dim mel-spectrogram in 50ms frame length and 12.5ms frame shift. We adopt the WeNet U2++ model~\cite{DBLP:conf/interspeech/YaoWWZYYPCXL21} trained with 10,000 hours of data in WenetSpeech corpus~\cite{DBLP:conf/icassp/ZhangLGSYXXBCZW22} as our ASR model, and the Conformer based encoder output is used as the BN feature with dimension of 512. The BN feature is further interpolated to match the sequence length of the mel-spectrogram.

\vspace{-0.16cm}
\subsection{Comparison Models}

\vspace{-0.1cm} 
\begin{itemize}
\item  \textbf{Accent-FastSpeech:} As a baseline, we use high-quality unaccented data (DB1) and low-quality accented data (Kespeech) to train a FastSpeech~\cite{DBLP:conf/nips/RenRTQZZL19} model which consists of encoder, decoder, duration predictor and LR. There are 6 layers of FFT in encoder and decoder. The duration predictor consists of 2-layer 1D-convolutional network with ReLU activation. The LR aims to expand phoneme-level representation to frame-level representation. The encoder accepts accent ID and linguistic features as input, and the decoder accepts the output of the LR and speaker ID as input.

\vspace{-0.1cm}
\item  \textbf{Accent-Hieratron:} We modify the Hieratron~\cite{DBLP:conf/icassp/DaiCCTLXTWW22} model to support accent transfer. Specifically, we use Kespeech to train a T2BN model in Hieratron for each accent. The T2BN model adopts the model structure of Accent-FastSpeech. We use DB1 as the target speaker to train the BN2Mel model in Hieratron. The BN2Mel model consists of CBHG, auto-regressive module and postnet. In inference, the T2BN model generates the BN feature with a specific accent. The BN2Mel model takes the BN feature and speaker ID as input to generate the mel-spectrogram of the target speaker. In this way, we can realize the decoupling of accent and speaker timbre.

\vspace{-0.1cm}
\item  \textbf{AccentSpeech (Proposed):} The T2BN model adopts the model structure of Accent-FastSpeech. The hidden and filter dimensions of the T2BN are 192 and 768, respectively. The dimension of the autoregressive layer is 256 in the BN2Mel model. In inference, the T2BN model generates the BN feature without an accent, and the BN2BN model accepts the unaccented BN as input to generate the accented BN. The BN2Mel model takes the accented BN feature and speaker ID as input to generate the mel-spectrogram for the target speaker.
\end{itemize}

\vspace{-0.1cm}
We use the text front-end of Mandarin to extract transcription from the text and the force-alignment tool~\cite{povey2011kaldi} to extract phoneme-level duration for all datasets. We use HiFiGAN~V1~\cite{DBLP:conf/nips/KongKB20} as the vocoder, trained for 500k steps. MSE loss is used as the loss function of T2BN, BN2BN and BN2Mel. The above comparison models are trained for 200k steps. 
The batch size of all the models is 16. The initial learning rate of all the models is 2e-4. The Adam optimizer with $\beta_{1}$ = 0.9, $\beta_{2}$ = 0.98 and $\epsilon$ = $10^{-9}$ is used to train all the models.

\vspace{-0.20cm}
\subsection{Visualization on Generated Parallel Data}

\begin{figure}[t]
  \centering
  \includegraphics[width=0.75\linewidth]{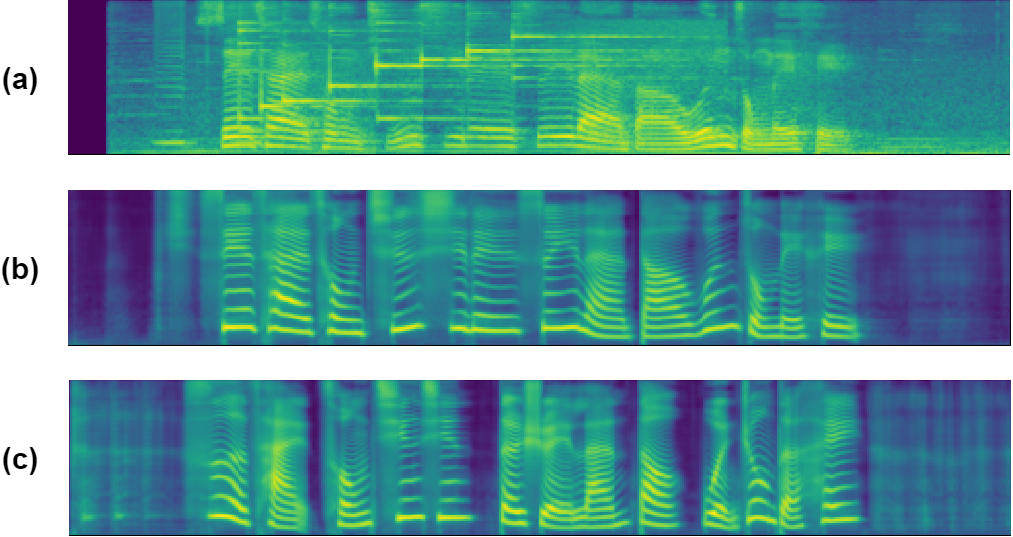}
  \caption{Visualization on generated parallel samples through data augmentation. }\vspace{-10pt}
  \label{fig:mel}
\end{figure}

\vspace{-6pt}
As described in Section~\ref{accent transfer}, the accent transfer model is trained by parallel data generated by data augmentation. Therefore, we visualize a pair of generated parallel samples, as shown in Figure~\ref{fig:mel}. Specifically, Figure~\ref{fig:mel} (a) shows the mel-spectrogram of a crowd-sourced recording from an accent speaker. We can clearly see the background noise (ringing tone) along with the speech collected during the audio recording process.
We extract the BN feature from the same audio in Figure~\ref{fig:mel} (a) and convert the BN to mel-spectrogram through the BN2Mel model trianed with the high-quality (DB1) data. The mel-spectrogram is shown in Figure~\ref{fig:mel} (b). We can see that most of the noise is removed. With the same transcript of the crowd-sourced audio recording in Figure~\ref{fig:mel} (a) as input, the high quality DB1 data trained T2BN + BN2Mel generates the corresponding mel-spectrogram shown in Figure~\ref{fig:mel} (c). Hence Figure~\ref{fig:mel} (a) and (c) are associated with the same linguistic content but difference accent. Thus the corresponding paired unaccented and accented BN features are used to train the BN2BN model for the purpose of accent transfer.

\vspace{-0.22cm} 
\subsection{Results and Analysis}

\vspace{-6pt}

We evaluate the speaker similarity, naturalness and accent similarity of different systems. The Accent-FastSpeech system can not converge because of the poor quality of the crowd-sourced data. Therefore we do not evaluate the results of Accent-FastSpeech here. The test set for TTS experimentation is composed of both short and long sentences without overlap on the training set. We randomly selected 10 utterances for each accent, and there are a total of 30 utterances for evaluation. Ten Chinese listeners are asked to evaluate the speaker similarity and naturalness of different systems. There are fifteen local accent listeners to evaluate the accent similarity, with five listeners for each accent (Chengdu, Xi'an and Zhengzhou). We suggest the readers to visit our demo page~$^1$.

\renewcommand{\thefootnote}{\arabic{footnote}}
\footnotetext[1]{\url{https://accentspeech.github.io/AccentSpeech/}}


\begin{figure}[t]
  \centering
  \includegraphics[width=0.85\linewidth]{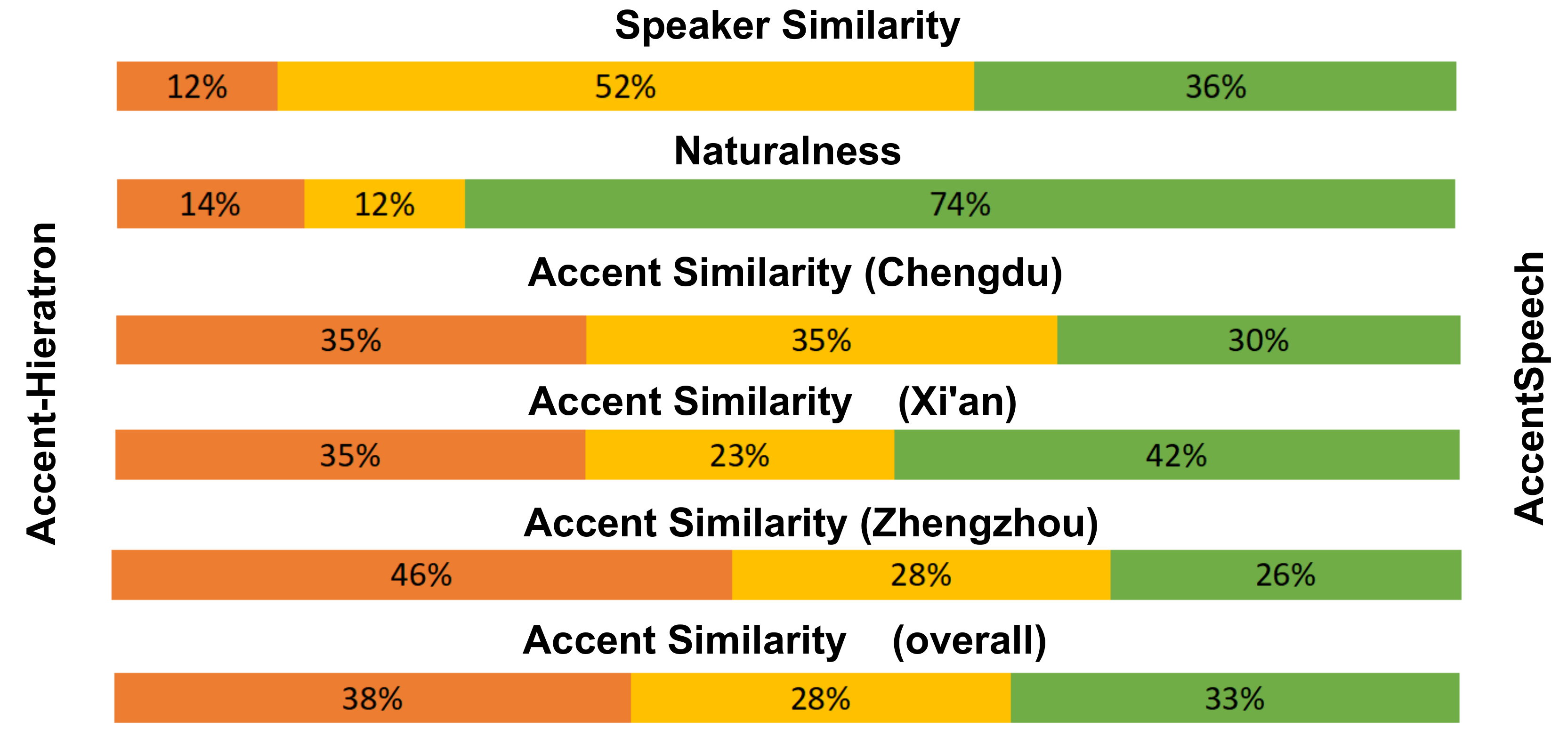}
  \caption{Preference test results of Accent-Hieratron vs AccentSpeech.}\vspace{-18pt}
  \label{fig:abtest}
  
\end{figure}

\textbf{Speaker similarity.} We conduct AB test on speaker similarity. An audio sample from the target speaker is served as the reference and the listeners are asked to judge which sample (from Accent-Hieratron or AccentSpeech) is more similar to the reference. From the first row in Figure~\ref{fig:abtest}, we can clearly see that AccentSpeech is more preferred by the listeners on speaker similarity. We also notice that there is also a large ratio on the no-preference option.

We calculate the cosine similarity on the generated samples to further verify the speaker similarity. Specifically, we train an ECAPA-TDNN model~\cite{DBLP:conf/interspeech/DesplanquesTD20} using 6000 hours of Mandarin speech from 18083 speakers to extract x-vectors. The cosine similarity to the target speaker audio is measured on 30 synthetic utterances. We can see from Table~\ref{tab:cos-sim and dur-mae} that the x-vector cosine similarity of the two systems are very close. When we remove the BN2BN accent transfer model from AccentSpeech, which indicates a pure BN-based target speaker TTS system without accent, the value of x-vector cosine similarity will become larger. This shows that the addition of the accent transfer module will result in a small decrease in objective measure of speaker similarity. But the previous subjective AB test shows that speaker similarity is still acceptable with a high preference.

\textbf{Naturalness.} In the AB test on naturalness, we ask the listeners to pay more attention to the general prosody such as rhythm and expressiveness of the audio. From the second row of Figure~\ref{fig:abtest}, we can see that the listeners have overwhelmingly voted to AccentSpeech.  Many listeners argue that the synthesized speech from Accent-Hieratron has poor rhythm, lacking natural prosody; on contrast, the synthesized samples by AccentSpeech are more natural on rhythm. This subjective conclusion is mainly attributed to the fact that Accent-Hieratron has inherited the poor prosody from the crowd-sourced accent speaker while AccentSpeech has learned the natural prosody from the professional speaker instead.

\begin{figure}[H]
  \centering
  \includegraphics[width=0.65\linewidth]{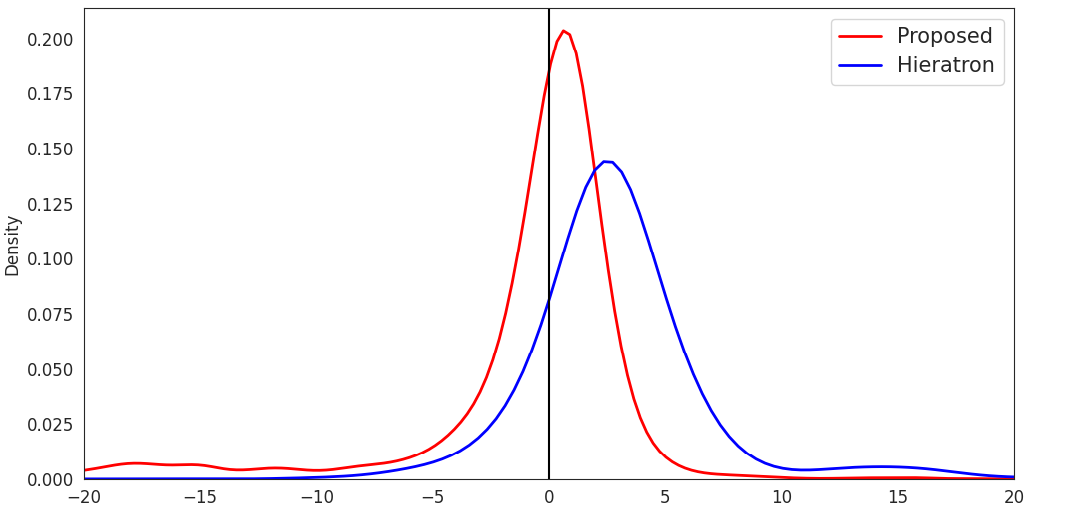}
  \caption{Duration deviation comparison for different systems.}
  \label{fig:dur_diff}\vspace{-15pt}
\end{figure}

Rhythm is largely reflected on the perceived duration of pronunciation units. Therefore we further visualize the results of duration prediction of different systems. The ground truth phoneme duration of each speech recording is obtained by a force-alignment tool~\cite{povey2011kaldi}. We randomly select 100 utterances from the test set of target speaker and count the phoneme-level duration difference between the predicted and the ground truth of the target speaker, the duration mean absolute error (MAE) is shown in Table~\ref{tab:cos-sim and dur-mae}, and the duration deviation is shown in Figure~\ref{fig:dur_diff}, where the horizontal axis represents the duration difference between the predicted and the ground truth, and the vertical axis indicates the frequency of the occurrence (density). The smaller the horizontal axis value is, the closer the predicted duration is to the duration of the ground truth speech of the target speaker. We can see that the predicted duration values in the result of the proposed AccentSpeech are more accurate to the target speaker. Again, this result shows that Accent-Hieratron has learned poor rhythm from the low-quality data while AccentSpeech has mitigated this problem and learned much better rhythm from the professional target speaker.

\begin{table}[t]
  \centering
   \caption{The speaker cosine similarity and duration MAE for difference systems.}
   
\vspace{-4pt}

   \resizebox{.95\columnwidth}{!}{
\begin{tabular}{@{}ccc@{}}
\toprule
\multirow{2}{*}{System} & \multirow{2}{*}{\begin{tabular}[c]{@{}c@{}}Speaker\\ cosine Similarity\end{tabular}} & \multirow{2}{*}{Duration MAE} \\
                        &                                    &                          \\ \midrule
Accent-Hieratron               & 0.697                              & 4.446                    \\
AccentSpeech            & 0.705                             & \textbf{2.607}                    \\
AccentSpeech~(w/o BN2BN)            & \textbf{0.760}                              &   \textbf{2.607}                    \\ \bottomrule
    \end{tabular}}
 \label{tab:cos-sim and dur-mae}\vspace{-15pt}
\end{table}

\vspace{-0.3cm}

\textbf{Accent similarity.} We finally access the accent similarity on the synthetic accent speech. To this end, we ask the accent listeners to pick the sample which is more similar to the target accent in terms of pronunciation, ignoring the naturalness on general prosody. The AB preference shown in Figure~\ref{fig:abtest} indicates that listeners slightly prefer AccentSpeech in Chengdu and Zhengzhou accents while they prefer Accent-Hieratron in Xi'an accent. According to the overall accent similarity preference in the last row of Figure 3, AccentSpeech and Accent-Hieratron are comparable in general. 

In summary, in the task of learning accent from crowd-sourced data for target speaker accent TTS, the proposed AccentSpeech approach has superior speaker similarity and naturalness as well as comparable accent similarity as compared with Accent-Hieratron, while naive control ID based FastSpeech approach fails.

\vspace{-0.30cm}
\section{Conclusions}
\vspace{-0.17cm}

In this paper, we propose AccentSpeech, a system built on neural bottleneck (BN) feature to learn accents from crow-sourced data for target speaker accent TTS. The system is composed of three specifically designed modules. The text-to-BN (T2BN) model and the BN-to-Mel-spectrum (BN2Mel) model are trained using the target speaker data to ensure the system to produce high quality voice with target speaker timbre and prosody. The BN2BN model in-between T2BN and BN2Mel is trained for transfering the accent from the crowd-sourced data to the target speaker, and the parallel unaccented and accented BN data for model training is obtained through data augmentation. AccentSpeech shows superior performance with good speaker similarity, naturalness and accent similarity in a Mandarin TTS accent transfer task.

\newpage

\bibliographystyle{IEEEtran}

\bibliography{mybib}

\end{document}